\documentclass{emulateapj}

\newcommand{\CII}{[\ion{C}{2}]}

\begin{document}

\title{A Detection of [\ion{C}{2}] line emission in the $z=4.7$ QSO BR1202-0725}
\author{Daisuke Iono\altaffilmark{1,2}, 
Min S. Yun\altaffilmark{3}, 
Martin Elvis\altaffilmark{2}, 
Alison B. Peck\altaffilmark{2}, 
Paul T. P. Ho\altaffilmark{2,4}, 
David J. Wilner\altaffilmark{2}, 
Todd R. Hunter\altaffilmark{2}, 
Satoki Matsushita\altaffilmark{4}, 
Sebastien Muller\altaffilmark{4} 
}
\altaffiltext{1}{National Astronomical Observatory of Japan, 2-21-1 Osawa, Mitaka, 181-8588 Tokyo, Japan; d.iono@nao.ac.jp}
\altaffiltext{2}{Harvard-Smithsonian Center for Astrophysics, 60 Garden Street, Cambridge, MA 02138}
\altaffiltext{3}{Department of Astronomy, University of Massachusetts, Amherst, MA 01003}
\altaffiltext{4}{Academia Sinica Institute of Astronomy and Astrophysics, P.O. Box 23-141, Taipei 106, Taiwan, R.O.C.}

\begin{abstract}

We present $\sim3''$ resolution imaging of the $z=4.7$ QSO BR1202-0725 at
900~$\micron$ from the Submillimeter Array.
The two submillimeter continuum components are clearly resolved from each 
other, and the positions are consistent with previous lower frequency images.
In addition, we detect [\ion{C}{2}] line emission from the northern component.
The ratio of [\ion{C}{2}] to far-infrared luminosity is 0.04\% for the
northern component, and an upper limit of $< 0.03\%$ is obtained for the
southern component. These ratios are similar to
the low values found in local ultraluminous galaxies, indicating that the
excitation conditions are different from those found in local field galaxies.
X-ray emission is detected by {\it Chandra} 
from the southern component at L$_{0.5-2keV}=3\times10^{45}$~erg~s$^{-1}$,
and detected at 99.6\% confidence from the northern component 
at L$_{0.5-2keV}\sim$3$\times$10$^{44}$erg~s$^{-1}$,  
%both submillimeter continuum components, 
supporting the idea that BR1202-0725 is {\em a pair}
of interacting galaxies at $z=4.7$ that each harbor an active nucleus.

\end{abstract}


\section{Introduction}

The far infrared (FIR) fine structure lines of abundant species 
(e.g. [\ion{C}{1}], [\ion{C}{2}], [\ion{O}{1}]) 
are recognized as one of the most 
important coolants in the warm ISM.  
In particular, the emission from singly ionized carbon 
(i.e. [\ion{C}{2}], $\lambda_{rest} = 158 \micron$)
is known to trace 
warm ($\geq 200$ K) and dense ($n_{cr} = 3 \times 10^3$~cm$^{-3}$) 
photo-dissociation regions \citep[PDRs; ][]{kaufman99}, 
directly identifying the 
gas surrounding active star forming regions.  
The [\ion{C}{2}] line is the dominant coolant  
in the Galactic ISM, accounting for 0.1 -- 1 \% of the 
FIR luminosity \citep{hollenbach97}.  
The  brightness of the line makes it an attractive tracer of star 
formation in galaxies, and a relation holds between [\ion{C}{2}] luminosity 
and the massive star formation rate derived from H$\alpha$ emission 
for galaxies that have L$_{\rm FIR}$~$< 3 \times 10^{10}$~L$_{\odot}$ \citep{boselli02}.  
In the FIR luminous galaxies (ULIRGs), the [\ion{C}{2}]-to-FIR flux 
ratio is an order of magnitude smaller than in normal 
star forming galaxies \citep{luhman98,luhman03}.  
The reason for the deficiency in  the [\ion{C}{2}] line luminosity is not 
well understood, but several possible scenarios have been offered
 \citep{luhman98, malhotra01,pierini03, luhman03}.

Despite the large line luminosity expected, detecting [\ion{C}{2}] line 
emission from luminous high redshift sources using a single-dish 
submillimeter telescope has proved difficult
\citep[e.g.][]{isaak94,vanderwerf98,bollato04,marsden05}. 
The low [\ion{C}{2}]-to-FIR ratio suggests that the physical properties 
in high-$z$ sources resemble those of the local ULIRG population
\citep[e.g.][]{blain02}.  
Detection of the \CII\ line in the $z=6.4$ QSO J1148+5251 has been
reported recently by \citet[][]{maiolino05}.
Here we report a first detection of the [\ion{C}{2}] line
emission from the $z=4.7$ hyper-luminous ($\sim 10^{13}$~L$_{\odot}$) 
QSO BR1202-0725
observed using the Submillimeter Array \citep[SMA; ][]{ho04}\footnote{The 
Submillimeter 
Array is a joint project between the Smithsonian Astrophysical Observatory 
and the Academia Sinica Institute of Astronomy and Astrophysics, and is 
funded by the Smithsonian Institution and the Academia Sinica.}.

BR1202-0725 is an optically bright QSO ($M_B = -28.5$)
with a $2\farcs3$ Lyman~$\alpha$ extension toward the northwest 
\citep{hu96, petitjean96, fontana98, ohyama04}.  
Radio continuum emission was detected at 1.4, 4.9 and 43~GHz 
\citep{yun00,carilli02}.  The emission at 1.4~GHz was 
resolved into two components; one at the location of the QSO and
the other $\sim4''$ northwest.  This apparent companion is also 
seen in millimeter continuum emission \citep{omont96,guilloteau99}.
Abundant CO emission  (M$_{H_2} \geq 10^{10}$~M$_{\odot}$) 
was detected toward both the optical
QSO and the northwestern companion \citep{omont96,ohta96,
guilloteau01,carilli02}.   The CO-derived redshifts are
$z=4.6947$ and 4.6916 for the QSO and the 
northwestern companion, respectively \citep{omont96}.

We adopt $H_0 = 70$~km~s$^{-1}$~Mpc$^{-1}$,
 $\Omega_m$ = 0.3, $\Omega_{\Lambda}$ = 0.7, which gives
$D_L = 43.27$~Gpc ($1''$ = 6.5~kpc)
for the luminosity distance to BR1202-0725. 

\section{Observation and Data Reduction}

%Observations of BR1202-0725 
%were obtained on February 22, 2005 using 5 antennas, April 5, 2005 using 6 antennas,  March 21, 2006 using 7 antennas, and April 19, 2006 using 8 antennas in the compact configuration of the SMA. 
%The average system temperature (DSB) was 
%240~K, 310~K, 300~K, 250~K, for the first, second, third, and the fourth 
%track respectively.  The average phase RMS toward the 9.0~Jy 
%(in Spring 2005) quasar 3C279 was 5 and 15 degrees, and toward the 
%9.5~Jy (in Spring 2006) quasar 3C273 was 14 and 6 degrees, respectively.
A total of 5 tracks (Table~\ref{log}) were obtained toward BR1202-0725, where 
the total on-source integration time was 19.6~hours.
The SIS receivers were tuned to a redshifted [\ion{C}{2}] 
frequency of 333.969~GHz ($900~\micron$) (LSB). 
The SMA correlator had 2~GHz total bandwidth with 0.8~MHz 
(0.7~km~s$^{-1}$) spectral resolution. 
The data were calibrated using the IDL based SMA calibration tool MIR. 
The quasars 3C279 and 3C273 were observed to calibrate the antenna 
based time dependent gain, and Callisto, 3C279, 3C273, and 1924-292 were used
to calibrate the bandpass.  The flux scaling is derived assuming
Callisto has a total flux density of 15 Jy at 334 GHz, and we estimate a 20\% 
uncertainty in the derived fluxes.

Imaging and analysis of the visibility data were carried out using MIRIAD. 
%Maximum sensitivity was achieved by adopting natural weighting, 
%which gave a synthesized beam size of $3\farcs6  \times 3\farcs3$ 
%(P.A. = $77^\circ$). 
%In addition,  
The angular resolution achieved by adopting ROBUST = 2 
was $3\farcs4  \times 2\farcs7$ (P.A. = $-8^\circ$). 
%in order to resolve the two components in BR1202-0725.
The continuum image constructed using the line-free 
data has an RMS noise of 3.0~mJy.
Although the varying S/N across the the SMA bandpass makes 
it difficult to characterize the overall noise properties, 
the RMS noise in the [\ion{C}{2}] spectrum calculated using 
several line-free regions of the data cube is estimated to be 7.5~mJy.

\section{Results}

\begin{figure}
\plotone{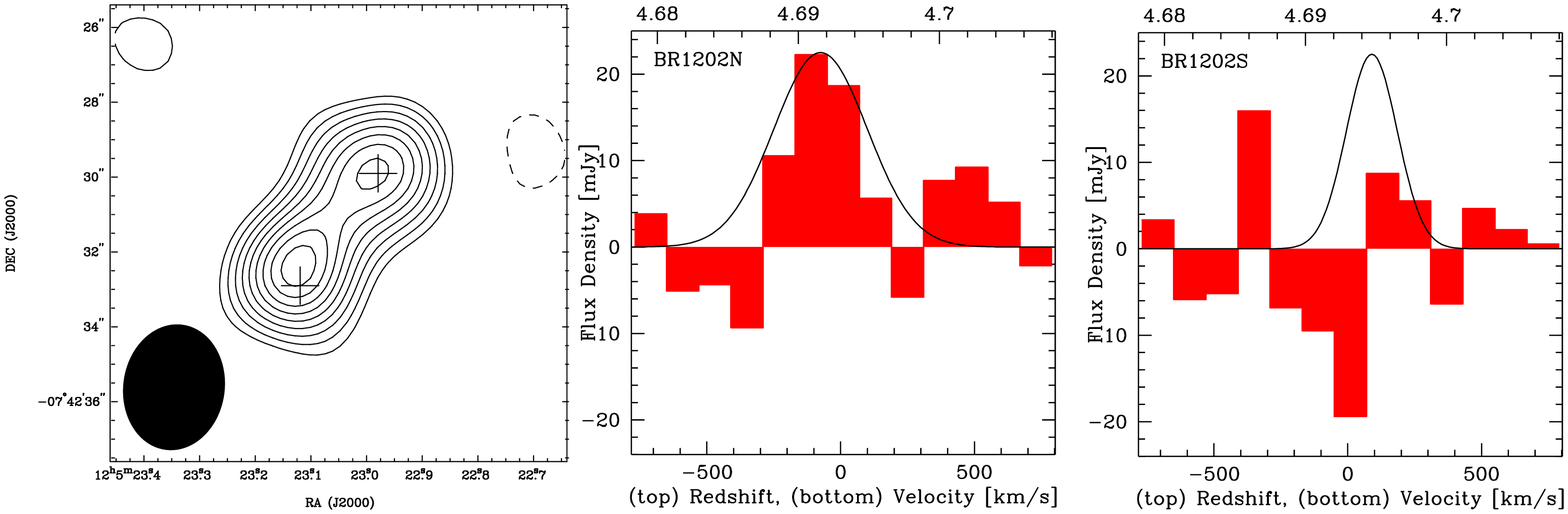}
\caption{(\textit{Left}) $900\micron$ continuum map of BR1202-0725 made by averaging line free channels in the LSB. The crosses mark the locations of the CO~(2--1) peaks from \citet{carilli02}.  The contour levels are -2,3,4,5,6,7,8,9,10,11$\sigma$~($1\sigma=3.0$~mJy~beam$^{-1}$).
The beam is shown in the lower left corner. 
The SMA line profile of BR1202N (\textit{middle}) and BR1202S 
(\textit{right}) obtained at the peak pixels using 120~km~s$^{-1}$ velocity 
averaging.  Velocity of 0~km~s$^{-1}$ corresponds to 334~GHz.  
The solid lines are schematic representations of the central velocities and linewidts of the CO~(5--4) line observed in BR1202N and BR1202S by \citet{omont96}, where the peaks have been scaled to match the [\ion{C}{2}] line intensity in BR1202N.  
}
\label{fig1}
\end{figure}

Figure~\ref{fig1}~($left$) presents the 900 $\micron$ continuum 
image of BR1202-0725.  
The coordinates of the two components are
$\alpha$~(J2000) = $12^h 05^m 22.98^s$ and 
$\delta$~(J2000) = $-7^{\circ} 42' 30\farcs0$ for BR1202N, and 
$\alpha$~(J2000) = $12^h 05^m 23.12^s$ and 
$\delta$~(J2000) = $-7^{\circ} 42' 32\farcs5$ for BR1202S
with uncertainties of $\pm 0\farcs2$.
The peak of the northern component (BR1202N) is coincident with  
the CO~(2--1) peak \citep{carilli02} within the $0\farcs2$ astrometric
accuracy. 
The peak of the southern component (BR1202S) is displaced to the north 
from the CO~(2--1) and the radio continuum peaks \citep{carilli02} 
by $\sim 0\farcs5$. The location of BR1202S is consistent with the optical/NIR QSO reported by \citet{hu96}.
The derived flux densities from the line-free region of the spectrum 
are $S_{900} = 27 \pm 4$~mJy 
(BR1202N) and $32 \pm 4$~mJy (BR1202S).
BR1202S emits $54\pm9\%$ of the total $900\micron$ flux, which is
slightly lower than the earlier observations at 1.3~mm in which 
$65\pm10\%$ of the total flux was found in BR1202S \citep{omont96}.

A spectrum at the peak of the $900\micron$ continuum in BR1202N
is shown in Figure~\ref{fig1}~(\textit{middle}). 
A similar spectrum toward BR1202S is also shown in 
Figure~\ref{fig1}~(\textit{right}). 
Continuum subtraction for both spectra was performed by fitting
a zeroth order baseline in the line-free channels of the visibilities.
The BR1202N spectrum shows a 22~mJy peak at $z \sim 4.691$ 
with a line profile that spans the redshift range 
of $z = 4.687$ -- 4.696.  No significant detection is seen in 
the BR1202S spectrum. 
The FWHM of the [\ion{C}{2}] line is $240 \pm 50$~km~s$^{-1}$,
which is narrower than the CO~(5--4) profile  (FWHM = $350\pm 60$~km~s$^{-1}$) 
obtained by \citet{omont96}.  
The formal significance of the [\ion{C}{2}] line detection over the
entire line-width shown in Figure~\ref{fig1} 
is $4\sigma$, assuming Gaussian statistics.  The true significance of this
feature is likely to be higher because noise is non-uniform across the
bandpass.  The excellent agreement between the [\ion{C}{2}] line
profile and the published CO~(5--4) line profile further supports the 
robustness of our [CII] line detection.
%#In order to investigate the exact spatial origin of the [\ion{C}{2}] 
%#line emission, an image using the emission in the velocity range -300 
%#to 200 km~s$^{-1}$ was made (Figure~\ref{fig1} ($c$)).  The result shows
%#that the emission primarily originates near BR1202N, 
%#with an extension toward BR1202S.  
%#Averaging line free channels of the same velocity 
%#width (500~km~s$^{-1}$) results in an image that is consistent 
%#with Figure~\ref{fig1} ($a$), 
%#demonstrating the robustness of the [\ion{C}{2}] line detection from BR1202N.  
By integrating the emission in the velocity range -230 to 130 km~s$^{-1}$,
we estimate the [\ion{C}{2}] line flux of 
$S_{\rm [CII]} = (6.8\pm1.1)$~Jy~km~s$^{-1}$
which translates to a [\ion{C}{2}] luminosity of 
$\rm L_{[CII]} = (4.5\pm0.7) \times 10^{9} L_{\odot}$.  
Using a FIR luminosity of 
$\rm L_{FIR} = (1.2\pm0.2) \times 10^{13} L_{\odot}$ (see \S4.2),
we obtain $\rm L_{[CII]}/L_{FIR} = (3.8\pm0.9) \times 10^{-4}$~(0.04\%)
in BR1202N.  In addition, assuming $\Delta v = 500$~km~s$^{-1}$ and 
$\rm L_{FIR} = 2.6 \times 10^{13} L_{\odot}$ (see \S4.2), we obtain a 
$3\sigma$ upper limit of $\rm L_{[CII]}/L_{FIR} < 2.8 \times 10^{-4}$~(0.03\%)
in BR1202S.

\section{Discussion}

\subsection{Properties of the [\ion{C}{2}] Line Emission}

The exceedingly low $\rm L_{[CII]}/L_{FIR}$ ratios for both BR1202N
and BR1202S are
consistent with the similarly low ratios found in local 
ULIRGs where the average was found to be
$\rm L_{[CII]}/L_{FIR} = 0.04\%$ \citep{luhman98,luhman03}.  
This is in stark contrast to the values found in the local 
galaxies by \citet{malhotra01} in which
$\sim 70\%$ of the sources show $\rm L_{[CII]}/L_{FIR} > 0.2\%$, with the 
majority in the range $0.1 - 1\%$. 
Past studies have reported 
$\rm L_{[CII]}/L_{FIR} \sim 0.02\%$ for the $z=6.42$ 
QSO SDSS~J1148+5251 \citep{maiolino05}, 
and $\rm L_{[CII]}/L_{FIR} < 0.4\%$ in a $z=4.926$ 
QSO CL~1358+62 \citep{marsden05}.
Our new measurement suggests that the [\ion{C}{2}]-to-FIR ratio 
(0.04\%) in the high-$z$ hyper-luminous source 
BR1202-0725 is comparable to local ULIRGs.
Table~\ref{ratios} summarizes these results.

The integrated [\ion{C}{2}] line intensity is known to be a tracer of 
star formation activity in normal galaxies.
We derive the star formation rate (SFR) of 2900~$\rm M_{\odot}~yr^{-1}$ 
using the calibration of \citet{maiolino05}.  
This is 1 -- 2 orders of magnitude higher than the
SFRs derived from optical recombination lines observed near BR1202N 
\citep[i.e. 10 -- 230 M$_\odot$~yr$^{-1}$;][]{hu96,petitjean96,
ohta00,ohyama04}, but consistent with the SFR derived from the SED fit to 
the millimeter/submillimeter continuum measurements 
(i.e. SFR = 2000 $\rm M_{\odot}~yr^{-1}$ in BR1202N; see \S 4.2).

Another important diagnostic of the physical properties 
is the [\ion{C}{2}]-CO flux ratio.
Early studies suggest the close spatial and kinematical association
between the [\ion{C}{2}] and CO~(1--0) emissions in nearby galaxies, 
suggesting that this ratio is a good tracer of star formation 
activity that is independent of the beam 
filling factor \citep{crawford85,stacey91}.
Assuming the luminosity ratio between the CO~(5--4) and CO~(1--0) lines 
in BR1202N is similar to that of 
the Cloverleaf \citep{barvainis94,barvainis97}, the CO~(1--0) luminosity is 
estimated to be L$_{\rm CO(1-0)} \sim 10^{6}$~L$_{\odot}$, yielding
 L$_{\rm [CII]}$/L$_{\rm CO(1-0)} \sim 4500$.
This value is highly uncertain since the CO~(5--4) -- CO~(1--0) ratio
may be different in BR1202-0725.  Nevertheless,
this is a factor of 2.5 larger than the average values of ULIRGs
\citep[i.e. L$_{\rm [CII]}$/L$_{\rm CO(1-0)} \sim$ 1700;][]{vanderwerf98},
but comparable to the average values of normal galaxies
\citep[i.e. L$_{\rm [CII]}$/L$_{\rm CO(1-0)}$ = 4400 -- 6300;][]{crawford85,
stacey91}.
The higher observed ratio may suggest a factor of few higher 
G$_0$/n ratio (where G$_0$ and n are the far UV flux and the cloud density
respectively) in BR1202N compared to the local ULIRGs 
\citep[see Figure~9 of][]{kaufman99}.

\subsection{Properties of the Continuum Emission}

\begin{figure}
\plotone{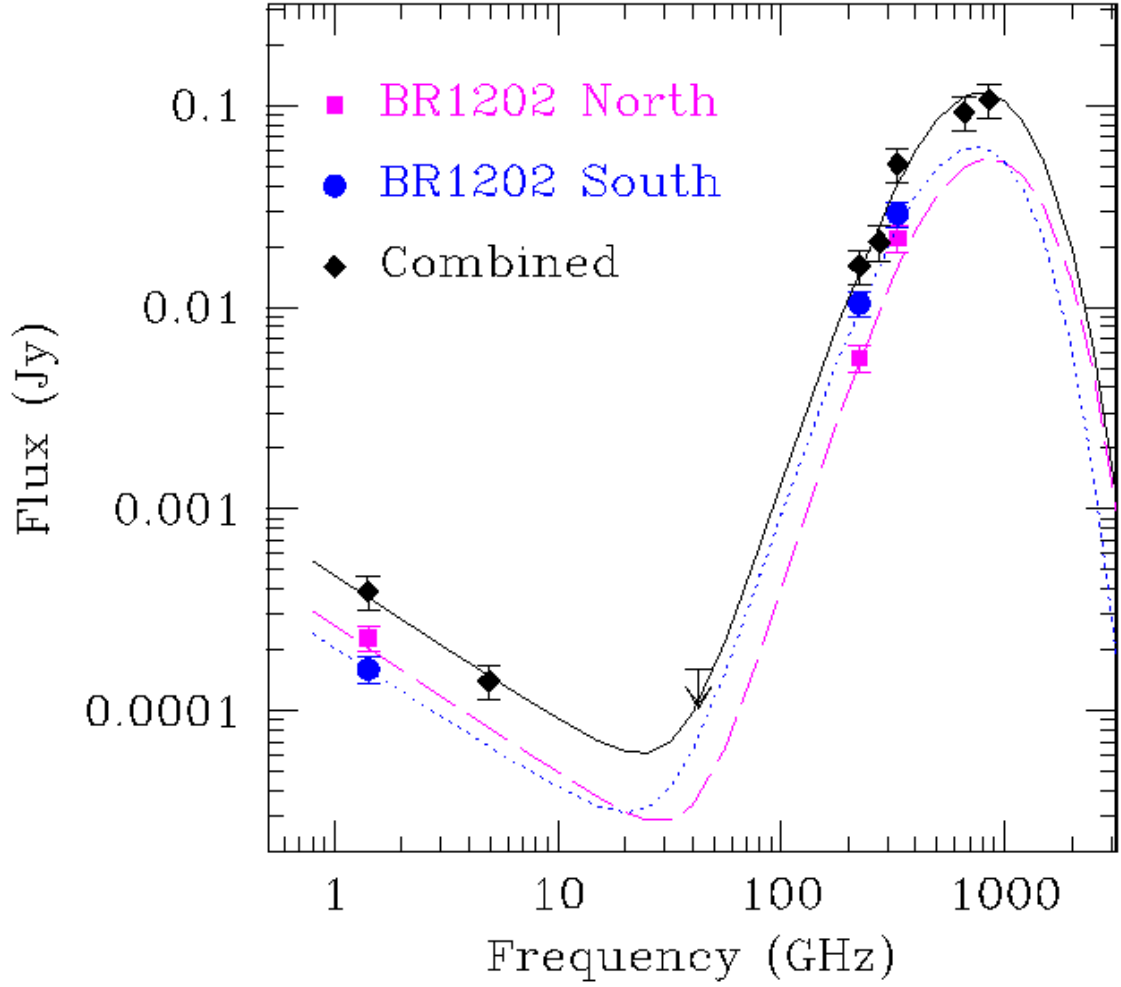}
\caption{The radio to FIR SED of BR1202-0725.  Two values are given when BR1202N and BR1202S are spatially resolved. The solid dark line represents the fit to the SED of the entire BR1202-0725 system, while the SEDs are not unique fits for BR1202N and BR1202S.  The parameters used for the 
SEDs are T$_d$ = 80K, $\alpha = 0.75$, 
$\beta = 1.75$, $f_{nth} = 6.0$ and SFR = 2000 M$_{\odot}~yr^{-1}$ 
for BR1202N and T$_d$ = 65K, $\alpha = 0.75$, 
$\beta = 1.50$, $f_{nth} = 2.0$ and SFR = 4500 M$_{\odot}~yr^{-1}$ 
for BR1202S, where $\alpha$, $\beta$, and $f_{nth}$ are the synchrotron spectral index, dust emissivity index, and the normalization factor to account for the non-thermal synchrotron emission \citep[see][for details]{yun02}. 
}
\label{fig2}
\end{figure}

The spatially resolved submillimeter images allow us to 
investigate the properties of the individual components in BR1202-0725.  
The radio to FIR SED of BR1202-0725 is shown in Figure~\ref{fig2}, where
the individual flux densities are given for BR1202N and BR1202S for the 1.4~GHz
\citep{carilli02}, 1.3~mm \citep{omont96}  and $900\micron$ observations (see figure caption for the adopted parameters).  The SEDs of the two sources
are analyzed using the model by \citet{yun02}, yielding 
L$_{\rm FIR} = 1.2 \times 10^{13}$ and $2.6 \times 10^{13}$~L$_{\odot}$, and 
SFR = 2000 and 4500 $\rm M_{\odot}~yr^{-1}$ 
for BR1202N and BR1202S respectively.
The discrepancy between the SFR derived by us and  
\citet{yun02}(2300 $\rm M_{\odot}~yr^{-1}$ for the entire BR1202-0725)
is due to the difference in the adopted cosmology.
Note that while the millimeter/submillimeter continuum flux densities of   
BR1202N and BR1202S are comparable (\S3), the radio
continuum flux density of BR1202N is 3 times larger.
This suggests a presence of a highly obscured AGN in BR1202N 
(see \S4.3), or a strong effect of a 
highly energetic jet from BR1202S \citep{klamer04}.
 
Assuming a dust absorption coefficient of $\kappa_d = 1.5$~m$^2$~kg$^{-1}$ at $160\micron$ \citep{hildebrand83,draine84}, 
the dust mass of the individual components is estimated to be 
M$_{\rm d}~(\rm BR1202N) = (5.4 \pm 0.8) \times 10^8$~M$_\odot$ and  
M$_{\rm d}~(\rm BR1202S)= (9.2 \pm 1.1) \times 10^8$~M$_\odot$.  
Using the H$_2$ mass derived from the CO~(5-4) emission \citep{omont96},
we find M$_{\rm H_2}$/M$_{\rm d} \sim 50 \pm 10$ for both components.  
This ratio is an order of magnitude lower than the mean value for 
LIRGs/ULIRGs
\citep[M$_{\rm H_2}$/M$_{\rm d} = 540 \pm 290$;][]{sanders91},
nearby spiral galaxies\footnote{The CO to H$_2$ conversion factor used by
\citet{omont96}, \citet{sanders91} and \citet{devereux90} are all
consistent to within 20\%.
In addition, the rest frame $160\micron$ emission for BR1202-0725 traces
the same warm dust (and hence the same M$_d$) 
as measured by {\it IRAS} \citep{yun06}.}
\citep[M$_{\rm H_2}$/M$_{\rm d} \sim 500$; see Fig. 2b of ][]{devereux90} 
and a sample of {\it Spitzer} 160 $\micron$ selected galaxies \citep[M$_{\rm H_2}$/M$_{\rm d} = 330$;][]{yun06}.
These gas-to-dust ratios are highly uncertain because dust emissivity, 
temperature, and CO-to-H$_2$ conversion factor are all poorly constrained.
If the low ratio for BR1202-0725 is correct, however, 
it may indicate that the H$_2$ mass is significantly underestimated in 
BR1202-0725, and/or the warm component of dust dominates 
the far infrared emission, suggesting that the entire 
ISM is involved in the starburst activity.

\subsection{BR1202-0725: A Colliding Galaxy System at $z=4.7$}
Our analysis of previously unpublished archival {\em Chandra} data 
(ObsID 3025) clearly shows that BR1202S is an X-ray source 
and gives 26.4 net counts in 9.64~ksec.
The derived flux from BR1202S is 
$f_{0.5-2keV} = 6\times10^{-15}$~erg~cm$^{-2}$~s$^{-1}$  which 
yields a luminosity of
L$_{0.5-2keV}=3\times10^{45}$~erg~s$^{-1}$
\citep[assuming $\Gamma$ = 2.0 and Galactic foreground 
N$_H$ = $3.35\times10^{20}$cm$^{-2}$;][]{stark92}.
X-ray emission is also seen within $1''$ of the position of BR1202N. 
With just 2 counts in a single pixel, this detection seems marginal, 
but the local background of 0.03 counts/pixel implies that
the likelihood of a random occurrence is only 0.04\%.
The derived flux from BR1202N is $f_{0.5-2keV} =
4.8^{+6.5}_{-4.3}\times$10$^{-16}$erg~cm$^{-2}$s$^{-1}$ 
\citep[95\% confidence;][]{regener51}, implying
L$_{0.5-2keV}\sim$3$\times$10$^{44}$erg~s$^{-1}$.

If the X-ray emission from BR1202N is real, its high X-ray luminosity
suggests that BR1202N also hosts a luminous AGN. 
In addition, our new [\ion{C}{2}] line detection, large dust emission, 
and the presence of abundant molecular gas suggest that BR1202N 
is forming stars at a high rate.  BR1202S and BR1202N thus
form a system of two massive galaxies
undergoing an interaction or a merger, separated by a projected 
distance of $\sim$25~kpc, observed when the universe was only 1.2 Gyrs old.
The presence of a highly obscured AGN (and an optically bright QSO) 
during a massive merger 
is consistent with the scenario predicted in recent galaxy collision 
simulations that investigate the evolution of a central 
black hole \citep[e.g.][]{hopkins05}.

A high degree of clustering near massive galaxies is expected in the  
theoretical models of large scale structure formation \citep[e.g.][]{white78}.
This is supported by observations such as the higher surface density of
submillimeter sources in the fields surrounding high-$z$ radio 
galaxies \citep{stevens03}.  
Although an earlier study by \citet{giallongo98} found little evidence
for clustering of galaxies near BR1202-0725 
with $R' \leq 25$, two massive galaxies in
proximity of each other makes BR1202-0725 a particularly interesting
test ground for the hierarchical scenarios, soon after the reionization.

\section{Summary}

We present a $3''$ resolution $900\micron$ continuum image and a  
detection of the redshifted [\ion{C}{2}] line emission from the $z=4.7$ 
QSO BR~1202-0725 obtained using the SMA. 
This is one of the first detections of the [\ion{C}{2}] line from a high redshift source.
The [\ion{C}{2}] line is associated with 
BR1202N.
The low [\ion{C}{2}]-FIR ratio of $\sim 3.8 \times 10^{-4}$ is similar 
to local ULIRGs. 
X-ray emission is clearly detected from BR1202S,  
and at 99.6\% confidence from BR1202N, suggesting
%in both submillimeter continuum sources, 
that BR1202-0725 is the first example of a pair of AGN hosts at $z \sim 4.7$.

The authors thank the anonymous referee for valuable comments, which have
improved this work significantly.

\clearpage

\clearpage

\clearpage

\begin{deluxetable}{lcccc}
\tabletypesize{\scriptsize}
\tablewidth{0pt}
\tablecaption{Summary of SMA Observations\label{log}}
\tablehead{
\colhead{Date} & \colhead{N$_{ants}$\tablenotemark{1}} & \colhead{$t_{\rm source}$\tablenotemark{2}} & \colhead{T$_{sys}$\tablenotemark{3}} & \colhead{RMS phase\tablenotemark{4}}  }
\startdata

02/22/2005 & 5 & 5.3 & 240 & 5 \\

04/05/2005 & 6 & 5.2 & 310 & 15 \\

03/21/2006 & 7 & 1.3 & 300 & 14 \\

04/19/2006 & 8 & 4.5 & 250 & 6 \\

05/11/2006 & 5 & 3.3 & 300 & 11  \\

\enddata
\tablenotetext{1}{Number of SMA antennas used.}
\tablenotetext{2}{On source integration time in hours.}
\tablenotetext{3}{The average DSB system temperature in Kelvins.}
\tablenotetext{4}{The average RMS phase (in degrees) taken over all baselines for each track.  These are calculated toward 3C279 for the first two tracks and 3C273 for the three tracks taken in 2006.}
\end{deluxetable}

\clearpage

\begin{deluxetable}{lccc}
\tabletypesize{\scriptsize}
\tablewidth{0pt}
\tablecaption{Summary of Luminosity Ratios\label{ratios}}
\tablehead{
\colhead{Source} &  \colhead{$\rm L_{FIR}$ (L$_\odot$)} & \colhead{$\rm L_{[CII]}/L_{FIR}$} & Ref.}
\startdata

Normal Galaxies  & $10^{7}$ -- $10^{11}$ & $0.1 - 1\%$&1 \\

ULIRGs  & $\sim 10^{12}$ & $0.01 - 0.1\%$ &2 \\

High-$z$ Sources \\

~~~BR1202N ($z=4.7$)  & $1.2 \times 10^{13}$ & $0.04\%$ & 3\\

~~~BR1202S ($z=4.7$)  & $2.6 \times 10^{13}$ & $<0.03\%$ & 3\\

~~~SDSS J1148+5251 ($z=6.4$) & $1.2 \times 10^{13}$ & $0.02\%$ & 4\\

~~~CL~1358+62 ($z=4.9$)  & $2.4 \times 10^{12}$ & $< 0.4\%$ & 5\\
\enddata
\tablerefs{(1) \citet{malhotra01}; (2) \citet{luhman03}; 
(3) this work; (4) \citet{maiolino05}; (5) \citet{marsden05}.}
\end{deluxetable}


\begin{thebibliography}{fun}
\bibitem[Barvainis et al.(1994)]{barvainis94} Barvainis, R., Tacconi, L., Antonucci, R., Alloin, D., \& Coleman, P. 1994, Nature, 371, 586
\bibitem[Barvainis et al.(1997)]{barvainis97} Barvainis, R., Maloney, P., Antonucci, R., \& Alloin, D., 1997, ApJ, 484, 695
\bibitem[Blain et al.(2002)]{blain02} Blain, A. W., Smail, I., Ivison, R. J., Kneib, J. -P. \& Frayer, D. T. 2002, PhR, 369, 111
\bibitem[Bolatto et al.(2004)]{bollato04} Bolatto, A. D., Di Francesco, J. \& Willott, C. J. 2004, ApJL, 606, L101 
\bibitem[Boselli et al.(2002)]{boselli02} Boselli, A., Gavazzi, G., Lequeux, J., \& Pierini, D. 2002, A\&A, 385, 454 
\bibitem[Carilli et al.(2002)]{carilli02} Carilli, C. L. et al. 2002, AJ, 123, 1838 
\bibitem[Crawford et al.(1985)]{crawford85} Crawford, M. K., Genzel, R., Townes, C. H., \& Watson, D. M. 1985, ApJ, 291, 75 
\bibitem[Devereux \& Young(1990)]{devereux90} Devereux, N. A. \& Young, J. S. 1990, ApJ, 359, 42
\bibitem[Draine \& Lee(1984)]{draine84} Draine, B. T. \& Lee, H. M. 1984, ApJ, 285, 89 
\bibitem[Fontana et al.(1998)]{fontana98} Fontana, A., D'Odorico, S., Giallongo, E., Cristiani, S., Monnet, G., \& Petitjean, P. 1998, AJ, 115, 1225
\bibitem[Giallongo et al.(1998)]{giallongo98} Giallongo, E., D'Odorico, S., Fontana, A., Cristiani, S., Egami, E., Hu, E., \& McMahon, R. G., AJ, 115, 2169
\bibitem[Guilloteau et al.(1999)]{guilloteau99} Guilloteau, S., Omont, A., Cox, P., McMahon, R. G. \& Petitjean, P. 1999, A\&A, 349, 363 
\bibitem[Guilloteau(2001)]{guilloteau01} Guilloteau, S. 2001, in ASP Conf. Ser. 235, Science with the Atacama Large Millimeter Array, ed. A. Wootten(San Francisco:ASP) 
\bibitem[Hildebrand(1983)]{hildebrand83} Hildebrand, R. H. 1983, QJRAS, 24, 267
\bibitem[Ho, Moran \& Lo(2004)]{ho04} Ho, P. T. P., Moran, J. M. \& Lo, Kwok Yung 2004, ApJ, 616, 1 
\bibitem[Hopkins et al.(2005)]{hopkins05} Hopkins, P. F., Hernquist, L., Cox, T. J., Di Matteo, T., Martini, P., Robertson, B., \& Springel, V. 2005, ApJ, 630, 705
\bibitem[Hu, McMahon \& Egami(1996)]{hu96} Hu, E. M., McMahon, R. G. \& Egami, E. 1996, ApJL, 459, L53 
\bibitem[Hollenbach \& Tielens(1997)]{hollenbach97} Hollenbach, D. J. \& Tielens, G. G. M., 1997, ARAA, 35, 179 
\bibitem[Isaak et al.(1994)]{isaak94} Isaak, K. G., McMahon, R. G., Hills, R. E. \& Withington, S. 1994, MNRAS, 269, L28 
\bibitem[Kaufman et al.(1999)]{kaufman99} Kaufman, M. J., Wolfire, M. G., Hollenbach, D. J. \& Luhman, M. L., 1999, ApJ, 527, 795
\bibitem[Klamer et al.(2004)]{klamer04} Klamer, I. J., Ekers, R. D., Sadler, E. M. \& Hunstead, R. W. 2004, ApJL, 612, L97
\bibitem[Luhman et al.(1998)]{luhman98} Luhman, M. L. et al. 1998, ApJL, 504, L11  
\bibitem[Luhman et al.(2003)]{luhman03} Luhman, M. L. et al. 2003, ApJ, 594, 758\bibitem[Maiolino et al.(2005)]{maiolino05} Maiolino, R., Cox, P., Caselli, P., Beelen, A., Bertoldi, F. et al. 2005, A\&A, 480, L51
\bibitem[Marsden et al.(2005)]{marsden05} Marsden, G., Borys, C., Chapman, S. C., Halpern, M. \& Scott, D. 2005, MNRAS, 359, 43 
\bibitem[Malhotra et al.(2001)]{malhotra01} Malhotra et al. 2001, ApJ, 561, 766 
\bibitem[Ohyama, Taniguchi \& Shioya(2004)]{ohyama04} Ohyama, Y., Taniguchi, Y. \& Shioya, Y. 2004, AJ, 128, 2704 
\bibitem[Ohta et al.(1996)]{ohta96} Ohta, K., Yamada, T., Nakanishi, K., Kohno, K., Akiyama, M. \& Kawabe, R. 1996, Nature, 382, 426 
\bibitem[Ohta et al.(2000)]{ohta00} Ohta, K. et al. 2000, PASJ, 52, 557 
\bibitem[Omont et al.(1996)]{omont96} Omont, A., Petitjean, P., Guilloteau, S., McMahon, R. G., Solomon, P. M. \& Pecontal, E. 1996, Nature, 382, 428 
\bibitem[Petitjean et al.(1996)]{petitjean96} Petitjean, P., Pecontal, E., Valls-Gabaud, D. \& Charlot, S. 1996, Nature, 380, 411 
\bibitem[Pierini, Leech \& V\"olk(2003)]{pierini03} Pierini, D., Leech, K. J., \& V\"olk, H. J. 2003, A\&A, 397, 871
\bibitem[Regener(1951)]{regener51} Regener, V. H. 1951, PhRv, 84, 161
\bibitem[Sanders, Scoville \& Soifer(1991)]{sanders91} Sanders, D. B., Scoville, N. Z., \& Soifer, B. T. 1991, ApJ, 370, 158
\bibitem[Stacey et al.(1991)]{stacey91} Stacey, G. J., Geis, N., Genzel, R., Lugten, J. B., Poglitsch, A., Sternberg, A., \& Townes, C. H. 1991, ApJ, 373,444 
\bibitem[Stevens et al.(2003)]{stevens03} Stevens, J. A. et al. 2003, Nature, 425, 264
\bibitem[Stark et al.(1992)]{stark92} Stark, A. A., Gammie, C. F., Wilson, R. W., Bally, J., Linke, R. A., Heiles, C. \& Hurwitz, M. 1992, ApJS, 79, 77
\bibitem[van der Werf(1998)]{vanderwerf98} van der Werf, P. P. 1998, in ASP Conf. Ser. 156, Highly Redshifted Radio Lines, ed. C. L. Carilli, S. J. E. Radford, K. M. Menten, \& G. I. Langston (San Francisco: ASP), 91. 
\bibitem[White \& Rees(1978)]{white78} White, S. D. M. \& Rees, M. J. 1978, MNRAS, 183, 341
\bibitem[Yun et al.(2000)]{yun00} Yun, M. S., Carilli, C. L., Kawabe, R., Tutui, Y., Kohno, K. \& Ohta, K. 2000, ApJ, 528, 171 
\bibitem[Yun \& Carilli(2002)]{yun02} Yun, M. S., \& Carilli, C. L. 2002, ApJ, 568, 88 
\bibitem[Yun et al.(2006)]{yun06} Yun, M. S., et al. 2006, in prep.
\end{thebibliography}
\end{document}